\documentclass[twocolumn]{article}  
\usepackage{graphicx}

\begin{document}
\title{Measuring the gain dynamics in a conjugated polymer
film}
\author{
Steven~A. van den Berg$^{a}$, Vladimir~A.~Sautenkov$^{a,b}$, Gert~W.~'t~Hooft$^{a,c}$, and Eric~R.
Eliel$^{a}\footnote{electronic address: eliel@molphys.leidenuniv.nl}$
\\
$^a$~Huygens Laboratory, Leiden University, P.O.Box 9504, 2300 RA Leiden, The Netherlands \\
$^b$~Lebedev
Institute, Leninsky Prospect 53, Moscow 119991, Russia \\
$^c$~Philips Research Laboratories, Prof. Holstlaan 4, 5656 AA Eindhoven, The Netherlands} \date{}\maketitle

\begin{abstract}
We present a simple method for measuring the gain decay time in a
conjugated polymer film by optically exciting the film with two
mutually delayed ultrashort pump pulses. When the pump is set at
such a power level that amplified spontaneous emission marginally
develops along the polymer waveguide, the total output emitted
from its edge decays exponentially as a function of the interpulse
delay. The corresponding decay time represents the decay time of the
gain of the polymer material.
\end{abstract}

\noindent Conjugated polymers are believed to carry great promise for
application as gain material in organic solid-state lasers. They
cover the whole visible spectral range and offer the prospect of
being electrically driven. Recently, a first electrically driven
organic laser has been realized using a crystalline
material~\cite{Schon2000}, but lasing from conjugated polymers has
so far only been achieved in an optically pumped device. In the
first optically pumped thin-film polymer laser a planar
microcavity was employed~\cite{Tessler1996}. Since then various
other optically driven polymer lasers have been reported, using a variety of
cavity configurations~\cite{Kawabe1998,McGehee1998a,Riechel2000}.

In a planar microcavity, a polymer layer is sandwiched between
two highly reflecting mirrors. The laser light then propagates
perpendicular to the active layer, just as in a vertical-cavity
surface-emitting semiconductor laser. In this configuration the
gain length is limited by the thickness of the layer or the
absorption length of the pump light (which usually doesn't exceed
100~nm). Contrary, this limit doesn't occur for configurations in
which the generated field propagates along the polymer film,
acting as a waveguide. In this case the gain length is determined
by the dimensions of the excitation beam, which offers the
possibility to generate strong amplified spontaneous emission
(ASE) along the waveguide, even in the absence of any
feedback~\cite{McGehee1998b,Zenz1998}.

To characterize the suitability of a specific conjugated polymer
film for laser applications, the gain and loss and their
corresponding dynamics need to be measured. The pump-probe
technique has proven to be especially powerful here, providing a
tool to measure stimulated emission and photoinduced absorption on
a sub-picosecond time
scale~\cite{Graupner1996,Schwartz1997,Jeoung1999,Wegmann1999}.
In this technique, the polymer film is excited with a femtosecond
pump pulse and the resulting change in optical transmission
($\Delta T/T$) is measured as a function of time by probing the
excited region with a very weak, delayed, ultrashort probe pulse.
Usually the probe pulse has a very broad spectrum, allowing
spectrally resolved measurements of stimulated emission ($\Delta
T/T>0$) and photoinduced absorption ($\Delta T/T<0$). In addition
to yielding the actual value of $\Delta T/T$  at zero delay for
all probe wavelengths, this technique also provides data on the
variation of $\Delta T/T$ as a function of time. Specifically, one
can determine the decay time of the stimulated emission. In order
to achieve lasing in a given device, this decay time should at
least exceed the time that is needed for the laser field to build
up in this device. In solid polymer films $\Delta T/T$ usually
diminishes on a time scale  much shorter than the
photoluminescence lifetime of the polymer. Various mechanisms
which have been heavily discussed in literature lie at the root of
this fast
decay~\cite{Jeoung1999,Long1997,Frolov1997c,Gelinck1997}.
These include photo-oxidation~\cite{Denton1997b}, film
morphology~\cite{Nguyen2000} and pump
intensity~\cite{Wegmann1999}.

In this Letter we put forward a new and simple method for
measuring the decay time of the gain in thin solid films of a
conjugated polymer. We employ the waveguide configuration and
excite the film in a pencil-shaped region using {\em two} pump
pulses, with a variable interpulse delay. We measure the
time-integrated output of the optically excited stripe as a
function of interpulse delay. We directly extract the gain
decay time from these measurements.

\begin{figure}[ht]
\centerline{\includegraphics[width=7cm]{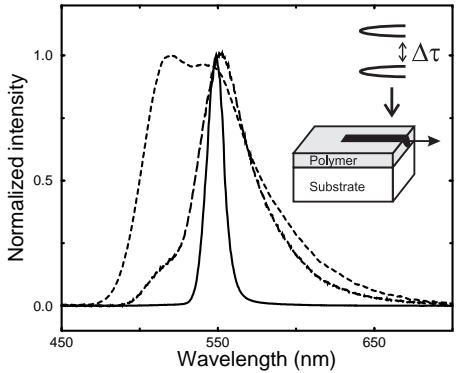}} \caption{Emission spectra measured for various pump energies.
The dashed curve shows the spontaneous-emission spectrum that is emitted from the surface of the waveguide for
very low pump intensity. The solid and long-dashed curves show the output emitted from the edge of the waveguide
at high pump energy ($>10$~nJ) and some intermediate pump energy, respectively.}
\label{ch.gaindynamics;Fig.spectrasample}
\end{figure}

Our sample is made by spin coating a heated and stirred solution
of a poly-($p$-phenylene vinylene) (PPV) copolymer in toluene (5~mg/ml) on top of a glass
substrate. The copolymer, referred to as HB1221, consists of four
different phenyl-PPVs (25\% each); the chemical
structure and synthesis route have been published in
Ref.~\cite{Spreitzer2001}. The substrate-film-air stack (with
refractive indices $\eta_{\mathrm{glass}}=1.46$, $\eta_{\mathrm
{poly}}\approx1.7$ and $\eta_{\mathrm{air}}=1$, respectively)
forms an asymmetrical waveguide with a thickness of
$\approx$100~nm, supporting the propagation of the fundamental
TE-mode~\cite{Zenz1998}.
 After spin coating the
substrate is broken in order to access the more homogeneous part
of the film in the middle of the substrate and to create a sample
with relatively sharp edges. We excite the film with two
ultrashort pump pulses ($\approx$ 130~fs, $\lambda=400$~nm, 1~kHz
repetition rate), one being delayed relative to the other. We
shape the excitation beam using lenses and a slit to make the
intensity distribution homogeneous. Subsequently it is focussed
into a stripe of $1\times0.02$~mm on the sample. We measure the
light emitted along the stripe. The setup is shown in the inset of
Fig.~\ref{ch.gaindynamics;Fig.spectrasample}.

\begin{figure}[ht]
\centerline{\includegraphics[width=7cm]{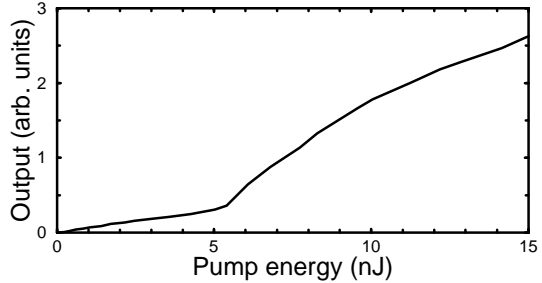}} \caption{The temporally and spectrally integrated output as
emitted from the edge of the waveguide measured as a function of the pump energy.}
\label{ch.gaindynamics;Fig.ase_inuit}
\end{figure}

First we provide some characteristics of the polymer waveguide
under single-pulse excitation. The spectrum of the emitted light
is shown in Fig.~\ref{ch.gaindynamics;Fig.spectrasample} for a
range of pump energies. At low pulse energy ($<1$~nJ) the broad
photoluminescence spectrum of the polymer film is observed (dashed
curve). When the pulse energy is increased a narrow peak develops
at $\lambda=548$~nm (long-dashed curve), which completely comes to
dominate the output at even higher pump energies (10~nJ, solid
curve). Another characteristic is the input-output curve of the
stripe as shown in Fig.~\ref{ch.gaindynamics;Fig.ase_inuit}.
It shows the temporally and spectrally integrated output
that is emitted from the edge of the polymer film. The signal is
measured with a slow photodiode in combination with a gated
integrator. Notwithstanding the absence of a feedback structure,
the curve shows a clear threshold around 5~nJ. In the vicinity of
this point the input-output curve is highly nonlinear. We will
employ this nonlinearity to obtain insight into the dynamics of
the gain.

The output spectral power on one side of the stripe with length
$l$ is given by:
\begin{equation}
P(\lambda,t)=A(\lambda)\frac{g(\lambda,t)}{g(\lambda,t)-\alpha}(e^{(g(\lambda,t)-\alpha)l}-1).
\label{ch.gaindynamics;Eq.stripeoutput}
\end{equation}
Here,  $g(\lambda,t)$ represents
the net optical gain, as induced by the pump pulse. 
The linear absorption in the waveguide is denoted by $\alpha$;
$A(\lambda)$ is related to the photoluminescence spectral power
per unit wavelength. The optical gain is associated with the
presence of primary excitons, which may be described as a
four-level system. The gain decay time is determined by radiative
and nonradiative decay of the exciton population.
Eq.~(\ref{ch.gaindynamics;Eq.stripeoutput}) has been derived from
the laser rate equations, assuming that the gain decays slowly as
compared to the single-pass transit time along the stripe
($\approx5$~ps); the latter time will be neglected henceforth.
Because we use ultrashort pump pulses, the gain coefficient is at
its maximum value at $t=0$ and subsequently diminishes. We assume
the gain decays via a single exponential.

In order to measure the decay time of the gain in our waveguide we
excite the polymer film at such a power level that there is very
little stimulated emission, the output being dominated by
spontaneous emission (e.g. at 4~nJ, see
Fig~\ref{ch.gaindynamics;Fig.ase_inuit}). After some delay we
reexcite the film with a second pump pulse, identical to the
first. We measure the time-integrated signal emitted by the stripe
upon this dual-pulse excitation as a function of the interpulse
delay. For excitation with two pump pulses of equal intensity,
having an interpulse delay $\Delta\tau$, the emission is given by
Eq.(\ref{ch.gaindynamics;Eq.stripeoutput}), with:

\begin{equation}  g_{\mathrm d}(\lambda,t)=\left\{
\begin{array}{lcl}
g_{\mathrm s}(\lambda,t) &\hspace{0.3cm}&t<\Delta\tau\\
g_{\mathrm s}(\lambda,t)+g_{\mathrm s}(\lambda,t-\Delta\tau) &\hspace{0.3cm}&t\ge\Delta\tau.\end{array} \right.
\end{equation}
Here $g_{\mathrm d}(\lambda,t)$ represents the double-pulse gain
coefficient, while the gain upon single-pulse excitation is
represented by $g_{\mathrm s}(\lambda,t)$. We now define
$U_{\mathrm s}$ and $U_{\mathrm d}(\Delta\tau)$ as the output
energies at the edge of the film for single and double pulse
excitation, respectively. The output energy is obtained by
integrating $P(\lambda,t)$ over both time and wavelength. When the
delay is much longer than the decay time of the gain, the
time-integrated double-pulse output simply equals twice the
single-pulse output ($U_{\mathrm d}(\Delta\tau)=2U_{\mathrm s}$).
However, when the delay is short, the double pulse output
$U_{\mathrm d}(\Delta\tau)$ is larger than $2U_{\mathrm s}$
because of the nonlinearity of
Eq.(\ref{ch.gaindynamics;Eq.stripeoutput}). This is most easily
seen at zero delay. The input-output curve immediately shows that
the output at 8~nJ pump energy is considerably larger than twice
the output at 4~nJ. It is easy to show that, for $0<(g_{\mathrm
s}(0)-\alpha)l< 1$, the excess output
\begin{equation}
U_{\mathrm{excess}}(\Delta\tau)=U_{\mathrm
d}(\Delta\tau)-2U_{\mathrm s}
\end{equation}
decays exponentially as a function of the interpulse delay
$\Delta\tau$, assuming $g_{\mathrm s}$ to decay exponentially. In
particular, it can be shown that the decay rate of $U_{\mathrm{excess}}$ directly yields the decay rate of the gain coefficient
$g_{\mathrm s}(t)$.

\begin{figure}[ht]
\centerline{\includegraphics[width=7cm]{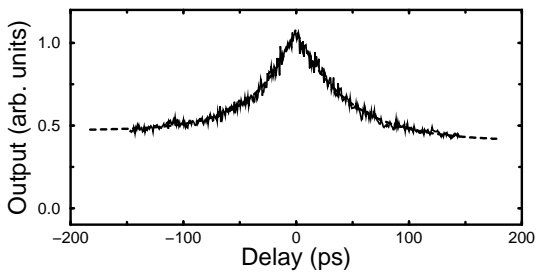}}
 \caption{The total edge-emission under double-pulse excitation
(4~nJ/pulse), measured as a function of the interpulse delay. The dashed lines represent single exponential fits.}
\label{ch.gaindynamics;Fig.ase4nJ}
\end{figure}
In Fig.~\ref{ch.gaindynamics;Fig.ase4nJ} we show the
time-integrated output that is emitted from the edge of our
substrate as a function of the interpulse delay at a single-pulse
excitation energy of 4~nJ/pulse ($\approx 20~\mu$J/cm$^2$).
As expected, the
curve is symmetrical with respect to zero delay. The fitted
exponentials (dashed curves), are in excellent agreement with the
experimental data, indicating that the excess output indeed decays
exponentially, the decay time being $\approx 40$~ps. The gain
coefficient is thus estimated to have a decay time of $\approx
40$~ps as well. The asymptotic values of the fits correspond to
twice the single pulse output, since in the long-delay limit the
two output pulses become independent.

\begin{figure}[ht]
\centerline{\includegraphics[width=7cm]{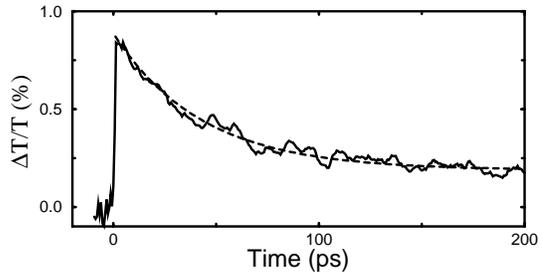}} \caption{Measurement of the decay time of the gain at 548~nm
using the pump-probe technique for a pump intensity of $\approx$ 80~$\mu$J/cm$^2$. The dashed line represents a
single exponential curve fit.} \label{ch.gaindynamics;Fig.pumpprobe}
\end{figure}
In order to check the validity of our approach we have also
applied the conventional pump-probe technique to our sample, again
using a low excitation density ($\approx 80~\mu$J/cm$^2$). The
probe wavelength has been set at $\lambda=548$~nm (gain maximum),
the same wavelength as selected by the gain medium when ASE
develops in the waveguide. The results are shown in
Fig.~\ref{ch.gaindynamics;Fig.pumpprobe}, representing a single
exponential decay of 43~ps, in excellent agreement with the result
obtained with the method introduced in this Letter.

Our technique thus provides a reliable tool to measure the
decay time of the gain in a thin film of light-emitting polymers.
The primary advantage of the technique introduced here is its
simplicity. No white-light generation is required and due to the
long gain length, the signals are large. This makes the method
well suited for studying systems that exhibit very small gain.
Moreover the gain decay time is automatically measured at the
wavelength at which the net gain is maximum and stimulated
emission will build up.

In conclusion, when pumping a polymer waveguide with two
excitation pulses at such a power level that the excitation energy
of a single pump pulse is just below the threshold value for ASE,
the total output upon double pulse excitation strongly depends on
the interpulse delay. We have shown that this behavior can be
exploited to measure the decay time of the gain in a very simple
way: the total output emitted from the edge of the waveguide
decays exponentially as a function of the interpulse delay with a
characteristic time that represents the decay time of the gain.

This work is part of the research programme of the Stichting voor Fundamenteel Onderzoek der Materie (FOM, financially supported by the Nederlandse Organisatie voor Wetenschappelijk Onderzoek (NWO)) and Philips Research.


\begin{thebibliography}{10}

\bibitem{Schon2000}
J.~H. Sch\"on, Ch. Kloc, A. Dodabalapur, and B. Batlogg, Science {\bf 289},
  599  (2000).

\bibitem{Tessler1996}
N. Tessler, G.~J. Denton, and R.~H. Friend, Nature {\bf 382},  695  (1996).

\bibitem{Kawabe1998}
Y. Kawabe, Ch. Spiegelberg, A. Sch\"ultzgen, M.~F. Nabor, B. Kippelen, E.~A.
  Mash, P.~M. Allemand, M. Kuwata-Gonokami, K. Takeda, and N. Peyghambarian,
  Appl. Phys. Lett. {\bf 72},  141  (1998).

\bibitem{McGehee1998a}
M.~D. McGehee, M.~A. D\'{\i}az-Garc\'{\i}a, F. Hide, R. Gupta, E.~K. Miller, D.
  Moses, and A.~J. Heeger, Appl. Phys. Lett. {\bf 72},  1536  (1998).

\bibitem{Riechel2000}
S. Riechel, U. Lemmer, J. Feldmann, T. Benstem, W. Kowalsky, U. Scherf, A.
  Gombert, and V. Wittwer, Appl. Phys. B. {\bf 71},  897  (2000).

\bibitem{McGehee1998b}
M.~D. McGehee, R. Gupta, S. Veenstra, E.~K. Miller, M.~A.
  D\'{\i}az-Garc\'{\i}a, and A.~J. Heeger, Phys. Rev. B {\bf 58},  7035
  (1998).

\bibitem{Zenz1998}
C. Zenz, W. Graupner, S. Tasch, and G. Leising, J. Appl. Phys. {\bf 84},  5445
  (1998).

\bibitem{Graupner1996}
W. Graupner, G. Leising, G. Lanzani, M. Nisoli, S.~De Silvestri, and U. Scherf,
  Phys. Rev. Lett. {\bf 76},  847  (1996).

\bibitem{Schwartz1997}
B.~J. Schwartz, F. Hide, M.~R. Andersson, and A.~J. Heeger, Chem. Phys. Lett.
  {\bf 265},  327  (1997).

\bibitem{Jeoung1999}
S.~C. Jeoung, Y.~H. Kim, D. Kim, J.~Y. Han, M.~S. Jang, J.~I. Lee, H.~K. Shim,
  C.~M. Kim, and C.~S. Yoon, Appl. Phys. Lett. {\bf 74},  212  (1999).

\bibitem{Wegmann1999}
G. Wegmann, B. Schweitzer, D. Hertel, H. Giessen, M. Oestreich, U. Scherf, K.
  M\"ullen, and R.~F. Mahrt, Chem. Phys. Lett. {\bf 312},  376  (1999).

\bibitem{Long1997}
X. Long, A. Malinowski, D.~D.~C. Bradley, M. Inbasekaran, and E.~P. Woo, Chem.
  Phys. Lett. {\bf 272},  6  (1997).

\bibitem{Frolov1997c}
S.~V. Frolov, M. Ozaki, W. Gellermann, M. Shkunov, Z.~V. Vardeny, and K.
  Yoshino, Synth. Met. {\bf 84},  473  (1997).

\bibitem{Gelinck1997}
G.~H. Gelinck, J.~M. Warman, M. Remmers, and D. Neher, Chem. Phys. Lett. {\bf
  265},  320  (1997).

\bibitem{Denton1997b}
G.~J. Denton, N. Tessler, N.~T. Harrison, and R.~H. Friend, Phys. Rev. Lett.
  {\bf 78},  733  (1997).

\bibitem{Nguyen2000}
T.-Q. Nguyen, I.~B. Martini, J. Liu, and B.~J. Schwartz, J. Phys. Chem. B {\bf
  104},  237  (2000).

\bibitem{Spreitzer2001}
H. Spreitzer, H. Becker, and P. St\"ossel, Substituted poly(arylene vinylenes),
  method for the production thereof and their use in electroluminescent
  devices, Patent nr. WO 01/34722, 2001.

\end{thebibliography}
\end{document}